\documentclass[aps,pre,groupedaddress,showkeys,showpacs]{revtex4}
\usepackage{epsf}
\usepackage{latexsym}
\usepackage{epsfig}
\usepackage{amsmath}
\usepackage{amsfonts}
\usepackage{amssymb}
\usepackage{graphicx}
\usepackage{yfonts}

\newcommand{\gb}[1]{\mbox{\boldmath{$#1$}}}
\newcommand{\prs}[1]{{\left(#1\right)}}
\newcommand{\chs}[1]{{\left\{#1\right\}}}
\newcommand{\col}[1]{{\left[#1\right]}}
\newcommand{\norm}{\mathcal{N}}
\newcommand{\prob}[1]{\mathcal{P}\prs{#1}}
\newcommand{\parf}{\mathcal{Z}_n}
\newcommand{\icn}[2]{{\left<{#1}_1\cdots {#1}_{#2}\right>}}
\newcommand{\pder}[2]{\frac{\partial #1}{\partial #2}}
\newcommand{\avg}[2]{{\left<#1\right>_{#2}}}

\newcommand{\hpi}{{\hat{\pi}}}
\newcommand{\hx}{{\hat{x}}}

\newcommand{\ts}{{\tilde{s}}}
\newcommand{\bx}{{\mathbf{x}}}
\newcommand{\bhx}{{\mathbf{\hx}}}
\newcommand{\bh}{{\mathbf{h}}}
\newcommand{\bK}{{\mathbf{K}}}
\newcommand{\bC}{{\mathbf{C}}}
\newcommand{\bv}{{\mathbf{v}}}
\newcommand{\tavg}[1]{{\avg{#1}{\bK,\bC,\Lambda}}}

\newcommand{\cut}[1]{{}}

\begin{document}

\begin{abstract}
Using methods of statistical physics, we study the average number
and kernel size of general sparse random matrices over $GF(q)$,
with a given connectivity profile, in the thermodynamical limit of
large matrices. We introduce a mapping of $GF(q)$ matrices onto
spin systems using the representation of the cyclic group of order
$q$ as the $q$-th complex roots of unity. This representation
facilitates the derivation of the average kernel size of random
matrices using the replica approach, under the replica symmetric
ansatz, resulting in saddle point equations for general
connectivity distributions. Numerical solutions are then obtained
for particular cases by population dynamics. Similar techniques
also allow us to obtain an expression for the exact and average
number of random matrices for any general connectivity profile. We
present numerical results for particular distributions.
\end{abstract}

\title[Kernel size and number of random matrices over $GF(q)$]
  {Typical kernel size and number of sparse random matrices over $GF(q)$ - a statistical
  physics approach}
\author{Alamino, R.C., Saad D.}
\affiliation{Aston University, Neural Computing Research Group, Birmingham B4 7ET, UK}

\maketitle

\section{Introduction}

Random matrices over $GF(q)$ are highly important in a number of
application areas ranging from biology to computer science and
telecommunication. One of the areas where they play a particularly
important role is coding theory~\cite{mceliece_book}. In
particular, linear codes are defined by the kernel of a
parity-check matrix, where each kernel vector is termed a codeword
and is associated with an original uncoded message vector by a
linear operation defined by a generator matrix. Well known
examples include the Hadamard codes, where properties of the
kernel and rank play an important role~\cite{Phelps05}, and
low-density parity-check codes (LDPC) which provide the best
performance to date in many noise regimes. Although the most
studied and applied case of LDPC codes is of binary codes over
$GF(2)$ there is a significant body of work, of both practical and
theoretical nature~\cite{daveyGFq}, on codes over more general
finite fields showing  an improvement in performance with respect
to the binary version. In particular, statistical physics based
analysis of LDPC codes over $GF(q)$ has been reported
in~\cite{Nakamura01}.

Low-density parity-check codes are based on random sparse
matrices, where the fraction of non-zero elements goes to zero as
the size of the matrix increases. In most studies of LDPC codes,
it is assumed that a parity-check matrix with $M$ rows
(parity-checks) and $N$ columns defines a code of rate $R=1-M/N$,
\emph{exactly}, which is equivalent to the assertion that the
number of vectors in the kernel (and therefore the number of
codewords) is exactly $q^{NR}$.

In addition to being an interesting applied problem, the
properties of these matrices are also of great interest from the
pure mathematical point of view and a number of papers has already
tried to answer related questions in different instances with a
mathematical rigorous approach~\cite{Cooper00,Blomer98, Feng07}.

In this contribution, we address two key properties of sparse
random matrices over $GF(q)$, namely the average dimension of
their kernel and the number of matrices for a given connectivity
profile, in the case of \emph{large matrices}. When the matrices
are large, keeping $N\rightarrow\infty$ with $M/N$ constant, the
problem can be mapped into a system of interacting ``spins'' and
the powerful machinery developed for the study of disordered spin
lattices in condensed matter physics can then be used, under some
assumptions, to obtain the required properties.

In order to keep this paper as self-contained as possible and make
it accessible to a broad readership, we provide in
section~\ref{section:Key} a brief introduction to $GF(q)$ matrices
and their properties, and to the basic statistical physics
methodology on which we have based our analysis. The usual
statistical physics approach to the analysis of LDPC codes over
the binary field $GF(2)$ is generalized in such a way that it can
be efficiently applied to any $GF(q)$ for a general connectivity
distribution of non-zero elements and then used to calculate the
average kernel dimension of sparse random matrices (SRM) in
section~\ref{section:Kernel}. Making use of techniques developed
in section~\ref{section:Kernel}, the number of matrices for a
given distribution of non-zero elements is then obtained for
various connectivity profiles, in section~\ref{section:NMatrices}.
Finally, we present a discussion of the obtained results in
section \ref{section:Conclusions}.

\section{Key Concepts}
\label{section:Key}

\subsection{$GF(q)$-Matrices}

A Galois field $GF(q)$ is a finite field with $q$ elements, i.e.,
a set of $q$ elements $\chs{0,...,q-1}$, which we symbolize by
integers for convenience, which is a commutative group under
addition $\oplus: GF(q)\rightarrow GF(q)$, defined as integer
addition \emph{mod $q$}, and with a monoid structure with respect
to a commutative multiplication operation $\otimes:
GF(q)\rightarrow GF(q)$. The field also includes the zero element
'0', mapping every other element to itself, and the identity '1';
an additional requirement is that the multiplication and addition
have the algebraic distributive property. This last requirement
restricts the number of elements to be $q=p^n$, where $p$ is a
prime number and $n$ an integer.

Entries in matrices over $GF(q)$ take values of numbers in the
field $GF(q)$, where the usual additions and multiplications
involved in their algebra are defined by the corresponding
operations over the Galois field. The kernel, or \emph{null
space}, of an $M\times N$ matrix $A$ is defined as the set of
vectors $\bv\in GF(q)^N$ such that $A\bv=0$, with all operations
in the field $GF(q)$. The kernel is a linear vector space and
therefore will have $q^{d(A)}$ vectors, where $d(A)$ is the kernel
dimension. The rank $r(A)$ of the matrix is obtained by the
rank-nullity theorem as $r(A)=N-d(A)$.

\subsection{Disordered Systems}

An interacting spin problem has two main elements: an interaction
defined between a number of spin units, collectively represented
by the vector $\gb{\sigma}=\prs{\sigma_1,...,\sigma_N}$,  in a
lattice and a local field which acts in each variable $\sigma_i$
separately. Disordered spin systems are systems where one or both
of these elements (interaction and field) is a random variable.
Usually, we are interested in the properties of very large
systems, where the number $N$ of spins becomes infinite, the
so-called thermodynamic limit.

The main properties of the system in the thermodynamic limit can
be derived from a key quantity, the free-energy $f$, which in
probabilistic terms corresponds to the cummulant generating
function. For disordered systems, in the cases where the
free-energy is self-averaging with respect to the disorder, we can
calculate this quantity as
\begin{equation}
  f = -\lim_{N\rightarrow\infty} \frac1{\beta N} \avg{\ln Z}{},
\end{equation}
where $\avg{\cdot}{}$ indicates the disorder average,
$Z=\sum_{\gb{\sigma}} e^{-\beta \mathcal{H(\gb{\sigma})}}$ is the
partition function and $\mathcal{H(\gb{\sigma})}$ is the
Hamiltonian of the system. Although the self-averaging property
should be rigorously investigated for each system, we will assume
it holds here.

In order to obtain the free-energy, a powerful technique is to
make use of the replica method, based on the identity
\begin{equation}
\label{eq:rReplicaIdentity}
   \col{\pder{}{n} \ln \avg{Z^n}{}}_{n=0} = \avg{\ln Z}{}~.
\end{equation}

Average quantities can then be calculated for integer $n$ and then
analytically continued to zero. The replica theory is commonly
used in the area of disordered systems and is known to provide
exact results in many regimes, which include both physical and
non-physical systems~\cite{Mezard87,Nishimori01}.

Many problems in computing and communication theory can be mapped
to spin systems. For instance, error-correcting codes, in
particular LDPC codes~\cite{ksRev} and hard computational problems
such as K-SAT~\cite{monasson2} and
graph-coloring~\cite{colouring,ColoringRSB}, can be mapped to
diluted spin systems with random $p$-spin interactions and local
fields. In the coding example, interactions are defined by the
parity-check constraints, while the local fields are induced by
the codeword and received message. In the statistical physics
treatment, for mathematical convenience, the message bits
$\chs{0,1}$ and '$\oplus$' operation are mapped onto spin values
$\chs{+1,-1}$ and multiplication using the mapping $x\rightarrow
(-1)^x$. Variables over a general finite field $GF(q)$, $q\neq2$
are typically first mapped onto a binary string and then, using
the spin values representation, transformed into a spin
system~\cite{Nakamura01}.

\section{Mapping $GF(q)$ Matrices into Spin Systems}
\label{section:Mapping}

The transformation
\begin{equation}
  \label{equation:map}
  \sigma(v) = (-1)^v,
\end{equation}
where $\sigma\in\chs{+1,-1}$ and $v\in\chs{0,1}$, is usually
employed to map the $GF(2)$ variables onto the binary
representation. This mapping can be generalized to any $GF(q)$
without an intermediate use of the binary field.

Under the operation $\oplus$, $GF(q)$ is homeomorphic to the
cyclic group of order $q$ and therefore has a representation as
the complex $q$-th roots of unity with the group homeomorphism
$\sigma:GF(q)\rightarrow \mathbb{C}$ given by
\begin{equation}
  \sigma(v) = \exp\prs{\frac{2\pi i}{q} v},
\end{equation}
such that for every $v_1,v_2\in GF(q)$
\begin{equation}
  \begin{split}
    \sigma(v_1\oplus v_2) &= \exp\col{\frac{2\pi i}{q} \prs{v_1\oplus v_2}}\\
                          &= \exp\col{\frac{2\pi i}{q} \prs{v_1 + v_2}}\\
                          &= \exp\prs{\frac{2\pi i}{q} v_1}\exp\prs{\frac{2\pi i}{q} v_2}\\
                          &= \sigma(v_1)\sigma(v_2).
  \end{split}
\end{equation}

This mapping has a clear geometric interpretation: $2\pi v/q$ is
an angle in the unit circle, such that each element of the Galois
field is being mapped onto a spin variable ``pointing'' in one of
$q$ possible angles. Using this mapping allows one to write the
null-space constraint for a general vector
$\bv=\prs{v^1,...,v^N}\in GF(q)^N$ as
\begin{equation}
  \label{equation:PCheck}
  \delta\prs{A\bv,0} =\prod_{i=1}^M \delta\col{\bigoplus_{j=1}^N \prs{A_{ij}\otimes v^j},0},
\end{equation}
with
\begin{equation}
  \label{equation:Delta_q}
  \delta\col{\bigoplus_{j=1}^N \prs{A_{ij}\otimes v^j},0}
                     = \frac1{\Delta(q)}\prod_{m=1}^{q-1} \chs{1-\exp\prs{-\frac{2\pi i}{q} m}
                        \prod_{j=1}^N \exp\col{\frac{2\pi i}{q} \prs{A_{ij}\otimes v^j}}},
\end{equation}
and
\begin{equation}
  \Delta(q) = \prod_{m=1}^{q-1} \col{1-\exp\prs{-\frac{2\pi i}{q} m}}.
\end{equation}

Using the properties of the complex roots of unity, the above
quantity $\Delta(q)$ can be shown (see appendix~\ref{appendix:Delta}) to be
real and equal to the order $q$ of the field.

Based on this representation, we can now define the
``magnetization'' of the original system in analogy with the spin
system as
\begin{equation}
  m = \frac1N\sum_{j=1}^N \sigma^j,
\end{equation}
and the overlap between two configurations $\gb{\sigma}$ and $\gb{\sigma}'$ as
\begin{equation}
  \rho = \frac1N\sum_{j=1}^N \sigma^j \sigma'^j,
\end{equation}
where we are now working with the spin variables already mapped to the the complex field $\mathbb{C}$ and therefore the
operations of multiplication and addition correspond to the usual ones in $\mathbb{C}$.

It turns out that this kind of representation allows a
factorization of the terms simplifying the equations and making
the replica calculations simpler, as we will see in the following.

\section{Average Properties of the Kernel}
\label{section:Kernel}

The dimension of the kernel of an $M\times N$ matrix $A$ over
$GF(q)$ can be written as $d(A)=\log_q \Omega$ where
\begin{equation}
\label{eq:Omega}
  \Omega=\sum_\bv \delta\prs{A\bv,0},
\end{equation}
is the number of vectors in the kernel, $\delta$ is the Kroenecker
delta and $\bv\in GF(q)^N$. Direct calculation of~$\Omega$
from equation (\ref{eq:Omega}) by straightforwardly substituting the Kroenecker delta by its
integral representation trivially reproduces the rank-nullity
theorem. This calculation is not presented here.

The quantity we are interested in here is the average kernel
dimension, more specifically, its density in the limit of large
matrices, defined as $Ts$ where
\begin{equation}
  s \equiv  \frac1T \lim_{N\rightarrow\infty} \frac{\avg{d(A)}{A}}{N}
     =      \lim_{N\rightarrow\infty} \frac1N \avg{\ln \Omega}{A},
\end{equation}
where $1/T=\ln q$ and $M/N\equiv \lambda$, with $\lambda$ a finite
positive constant. Using the replica identity
(\ref{eq:rReplicaIdentity}), we can write
\begin{equation}
  s = \lim_{N\rightarrow\infty} \col{\pder{}{n} \ln \avg{\Omega^n}{A}}_{n=0}.
\end{equation}

The randomly chosen sparse matrices $A$ have exactly $K_i$
non-zero elements in the $i$-th row with probability $\prob{\bK}$,
$\bK\equiv\prs{K_1,...,K_M}$, and $C_j$ elements in the $j$-th
column with probability $\prob{\bC}$,
$\bC\equiv\prs{C_1,...,C_N}$, obeying the constraint
$\Lambda\equiv\sum_i K_i = \sum_j C_j$, where $\Lambda$ is the
total number of non-zero elements of the matrix. The elements of
$A$ are sampled from the finite field $GF(q)$ with independent
equal probabilities $\prob{A_{ij}}$.

Let us define, for brevity of notation,
$\parf\equiv\avg{\Omega^n}{A}$. Although the calculations,
presented in appendix~\ref{appendix:Replicas}, are similar to related
calculations in~\cite{Alamino07,Tanaka03b}, we will use a
different approach which is conceptually clearer and has the
advantage of allowing later generalizations. In this approach, we
sum directly over all entries of the matrix instead of defining a
connectivity tensor as used elsewhere~\cite{Alamino07,Tanaka03b},

\begin{equation}
  \label{equation:parf}
  \begin{split}
    \parf &= \left<\frac1{\norm} \sum_\chs{A_{ij}} \col{\prod_{i,j} \prob{A_{ij}}}
             \col{\prod_{i=1}^M\delta\prs{\sum_{j=1}^N\chi\prs{A_{ij}},K_i}}
             \col{\prod_{j=1}^N\delta\prs{\sum_{i=1}^M\chi\prs{A_{ij}},C_j}}\right.\\
          &  \quad\times\left. \prod_{a=1}^n \col{\sum_{\bv_a}\delta\prs{A\bv_a,0}}
             \right>_{\bK,\bC,\Lambda},
  \end{split}
\end{equation}
where the average is over the probability distribution
$\prob{\bK,\bC,\Lambda}$ with $\chi\prs{A_{ij}}=0$ if $A_{ij}=0$
and 1 otherwise, and the normalization $\norm$ gives the number of
matrices which obey the constraints averaged over the
distributions of the entries. In this way, any type of constraint
on the matrix can be readily included in the calculation, which
could be rather cumbersome in other approaches, based on the
introduction of a connectivity tensor as the corresponding
constraints have to be written in terms of the tensor elements,
which can be extremely complicated.

We refer the reader to appendix~\ref{appendix:Replicas} for  details of the
calculations. Using the replica symmetric ansatz, which is shown
to be exact for this problem (see appendix~\ref{appendix:Proof_RS}) we
arrive at the following self-consistent saddle point equations
\begin{align}
  \label{equation:S_Point_x}
  \hpi(\hx) &= \frac1{\alpha\epsilon(\alpha)}\sum_{i=1}^M
               \avg{\frac{\alpha^\Lambda}{\Lambda!} K_i \delta\prs{\hx-\prod_{l=1}^{K_i-1}x_l}}{\bx,\bK,\bC,\Lambda},\\
  \label{equation:S_Point_hx}
  \pi(x)    &= \frac1{\alpha\epsilon(\alpha)}\sum_{j=1}^N\avg{\frac{\alpha^\Lambda}{\Lambda!} C_j
               \delta\prs{x\!-\!
               \frac{\prod_{l=1}^{C_j\!-\!1}\col{1\!+\!(q\!-\!1)\hx_l}\!-\!\prod_{l=1}^{C_j\!-\!1}\prs{1\!-\!\hx_l}}
               {\prod_{l=1}^{C_j\!-\!1}\col{1\!+\!(q\!-\!1)\hx_l}\!+\!(q\!-\!1)\prod_{l=1}^{C_j\!-\!1}\prs{1\!-\!\hx_l}}}}{\bhx,\bK,\bC,\Lambda},\\
  \label{equation:S_Point_Q}
  0         &= \tavg{\frac{\alpha^\Lambda}{\Lambda!}\prs{1-\frac{\Lambda}{\alpha}}},
\end{align}
with
\begin{equation}
  \epsilon(\alpha) = \tavg{\frac{\alpha^\Lambda}{\Lambda!}},
\end{equation}
and to the corresponding expression for $s$
\begin{equation}
  \begin{split}
     s &= -\lambda\ln q -\frac{\alpha}N \avg{\ln \col{1+(q-1)x\hx}}{x,\hx}\\
       &  \quad + \frac1{N\epsilon(\alpha)} \sum_i
          \tavg{\frac{\alpha^\Lambda}{\Lambda!} \avg{\ln \col{1+(q-1)\prod_{l=1}^{K_i} x_l}}{\bx}}\\
       &  \quad + \frac1{N\epsilon(\alpha)} \sum_j\tavg{\frac{\alpha^\Lambda}{\Lambda!}
          \avg{\ln\chs{\prod_{l=1}^{C_j}\col{1+(q-1)\hx_l}+(q-1)\prod_{l=1}^{C_j}\prs{1-\hx_l}}}{\bhx}}.
  \end{split}
\end{equation}

It must be noted that the above equations are only meaningful if
$\Lambda\propto N$. A striking property of the above equations is
that they are \emph{completely independent} of the specific
distribution of the individual elements of the matrix, depending
only on the distribution of $\bK$ and $\bC$ (and, obviously, of
$\Lambda$).

There exists two straightforward analytical solutions of the above
equations, namely, the paramagnetic one given by
\begin{equation}
  \hpi(\hx)=\delta(\hx),\,\pi(x)=\delta(x),
\end{equation}
and the ferromagnetic solution
\begin{equation}
  \hpi(\hx)=\delta(\hx-1),\,\pi(x)=\delta(x-1).
\end{equation}

When substituted in the above equations, the paramagnetic solution
gives the average kernel density as $Ts=1-\lambda=1-M/N$
independently of the order $q$ of the finite field used. In the
case of LDPC codes defined by such matrices, this corresponds to
random parity-check matrices that defines a code of rate
$R=1-\lambda$. The average rank density in this case is $\lambda$.
The ferromagnetic solution gives $Ts=0$ and the matrix is full
rank; which incidentally means that such matrices cannot be used
to define a parity-check code due to the lack of redundancy.

These quantities can be associated to analogous quantities in the
statistical mechanics framework. We start by associating the
average rank density with the free-energy $f$ and writing
\begin{equation}
\label{eq:free_energy}
  f \equiv \frac{\avg{r(A)}{A}}N = 1-Ts,
\end{equation}
which allows one to associate $s$ with the entropy and the
internal energy density being constrained to be $u=1$. Defining $\beta=1/T$,
equation~(\ref{eq:free_energy}) becomes
\begin{equation}
  \begin{split}
    \beta f &= 1-\frac1N\avg{\ln \sum_\bv \delta\prs{A\bv,0}}{A}\\
      &= \frac1N \avg{N\ln e^\beta-\ln \sum_\bv \delta\prs{A\bv,0}}{A}\\
      &= -\frac1N \avg{\ln \sum_\bv \delta\prs{A\bv,0}\,e^{-\beta N}}{A}\\
      &= -\frac1N \avg{\ln \sum_\bv e^{-\beta \mathcal{H}(\bv)}}{A},
  \end{split}
\end{equation}
where the Hamiltonian of the corresponding statistical mechanical
system is formally
\begin{equation}
  \mathcal{H}(\bv) \equiv N-\ln \delta\prs{A\bv,0}.
\end{equation}

We solved the saddle point equations by means of population
dynamics for three different cases, in all of which we keep $K$
fixed
\begin{enumerate}
  \item Regular matrices - $C$ and $K$ fixed;
  \item Fixed $K$ and $\bC$ drawn from a multinomial uniform probability
    \begin{equation}
      \prob{\bC}=\frac{(MK)!}{\prod_j C_j!} \frac1{N^{MK}};
    \end{equation}
  \item Fixed $K$ while $\bC$ values are drawn from a Poisson integer distribution
  of mean $\Lambda/N=\lambda K$, for each column separately, until the limit of $MK$
  non-zero elements is reached.
\end{enumerate}

\begin{figure}
\centering
\includegraphics[width=14cm]{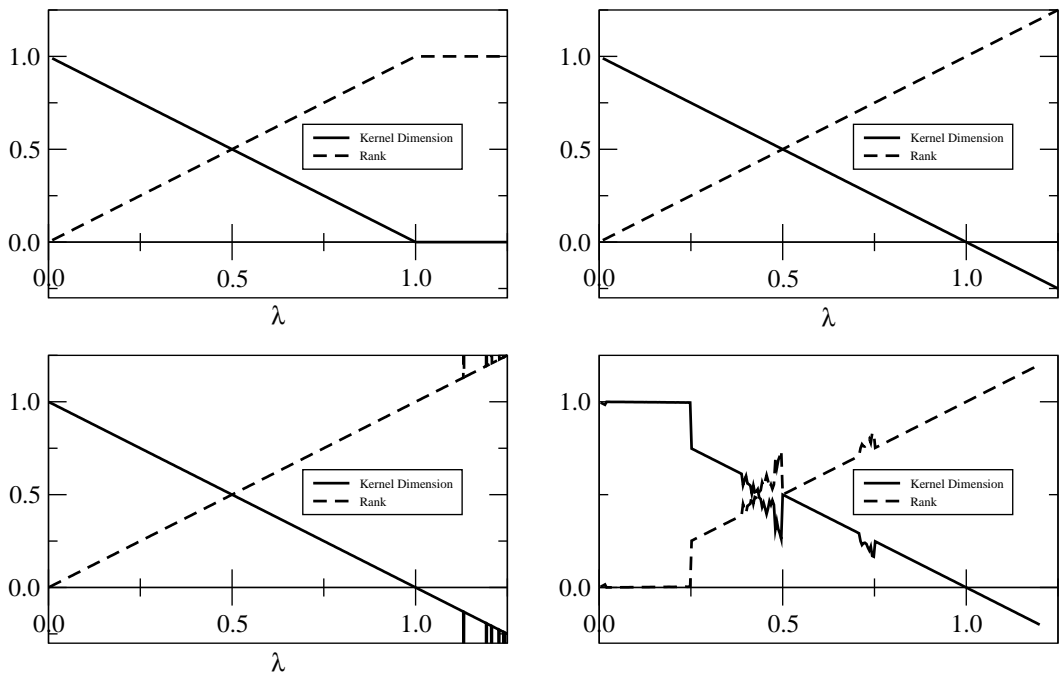}
\caption{Average kernel dimension density (continuous lines) and average rank density
(dashed lines) calculated as solutions to the replica symmetric saddle point equations.
The top left plot shows the thermodynamically favored solution (paramagnetic for
$0 \le \lambda \le 1$ and ferromagnetic for $\lambda > 1$). The top right shows
the regular case (i) for fixed $K$ and $C$. Cases (ii) and (iii) are presented
at the bottom left and right, respectively. Note that numerical instabilities occur for
specific $\lambda$ values.}
\label{figure:AKD}
\end{figure}

Results for the various cases are presented in Fig.~\ref{figure:AKD}. The top left
plot shows the theoretical thermodynamically dominant solutions (paramagnetic in the
range $0 \le \lambda \le 1$ and ferromagnetic for $\lambda > 1$) having the lower free energy.

The top right plot shows the results for the regular case (i). Solutions were obtained numerically
by iterating equations~(\ref{equation:S_Point_x}) and (\ref{equation:S_Point_hx}) 
for the case of $q=4$ and $K=200$; $C$ was varied from 2 to 250. Repeating the calculations for
different values of $q$ and $K$ have produced similar results. We see that the stable solution is always paramagnetic, 
but becomes unphysical at $\lambda=1$ once the entropy, and consequently the dimension of the kernel, become negative.

In the case of parity-check codes, this result means that the typical parity-check matrix defines a code of rate exactly 
$(N-M)/N$. This is assumed for any parity-check matrix in most calculations in the literature and is confirmed by our 
results to be true on average; however, it is important to point out that the result is true in the limit of large 
matrices and is likely to have finite size corrections which may affect practical applications.

Cases (ii) and (iii) are presented, respectively, at the bottom left and right of Fig.~\ref{figure:AKD}. Although these 
cases do not rigorously obey the constraint that each $C_j$ must be at most $M$, for large matrices and small values of 
$K$ (which is what happens in practice) $C_j$ is unlikely to
exceed this value. However, instabilities can and indeed occur for specific $\lambda$ values, presumably due to instances 
where $C_j$ takes higher values.

The  bottom left plot shows results for the case (ii), with $q=3$, $K=4$, $N=1000$ and $1\leq M \leq 1250$. Also in this 
case, the stable dominant solution is paramagnetic. Numerical instabilities, which disappear slowly with the increase in 
the number of fields and steps in the population dynamics, emerge in the unphysical region and are shown in the figure.

The behavior for case (iii) is a little more complex due to the nature of the distribution chosen. Using the average 
value $\lambda K$ for the variables $C_j$ implies that, as $\lambda$ varies, their average value also changes. The plot 
shown was obtained for $q=2$, $K=4$, $N=250$ and $1\leq M\leq 300$. There are clearly special points in this plot, which 
distinguish it from the previous cases. The first point separates $\lambda$ values which give rise to average connectivity
values lower/higher than 1 (left and right, respectively). Up to this point, the matrix has too many zero columns, pushing
the kernel size to cover the full space of vectors. The other two points are where numerical instabilities emerge. Further
calculations with different $K$ values indicate that these points appear around the extremes of the interval 
$2/K\leq\lambda\leq 3/K$. Inside this interval, the average value of the $C_j$'s equal to 2 (once we take it to be an 
integer). This value marks the percolation transition for binary matrices. 
Apart from these differences, the resulting curve seems to coincide with those obtained for the previous cases.

The solution of kernel size problem is mathematically equivalent to the solution
of LDPC in channels with infinite noise. As the solution in the latter is paramagnetic,
we are led to speculate that it is the dominant solution also here up to the point
where the quantity $s$, analogous to the entropy, becomes negative.
From this point and on the solution becomes ferromagnetic.
The numerical results seem to support this conjecture, although more careful calculations, varying all the parameters 
involved must be carried out to confirm this hypothesis more generally.

\section{Number of Matrices}
\label{section:NMatrices}

The number of $GF(q)$ matrices given a connectivity profile is of significant
interest within the discrete mathematics community. Exact results have been obtained
for the case of {\em finite} binary matrices~\cite{Wang98} in the form of a formula that
facilitates the calculation of their precise number. In this paper we will
analyze the case of large $GF(q)$ matrices and provide an expression for both their exact and
average number. Given the precise number of non-zero elements per row
$\bK=\prs{K_1,...,K_M}$ and per column $\bC=\prs{C_1,...,C_N}$, one
can write the number of matrices as
\begin{equation}
    N_A = \sum_\chs{A_{ij}}
          \col{\prod_{i=1}^M\delta\prs{\sum_{j=1}^N\chi\prs{A_{ij}},K_i}}
          \col{\prod_{j=1}^N\delta\prs{\sum_{i=1}^M\chi\prs{A_{ij}},C_j}}.
\end{equation}

Note that we are using the summation directly over the entries of the matrix instead
of the introduction of a connectivity tensor.
In this way, the calculations are similar to the ones for obtaining the kernel dimension
with the details given in~\ref{appendix:Matrices}. The final result is
\begin{equation}
  N_A= (q-1)^\Lambda \frac{\Lambda!}{\prod_i K_i! \prod_j C_j!}.
\end{equation}

Note that the component on the right represents the number of binary matrices
with the given non-zero elements
profile. The factor $(q-1)^\Lambda$ is the multiplicity of the non-zero entries which
can have any non-zero value in the
Galois field.

If we consider a distribution $\prob{\bK,\bC,\Lambda}$, we can look at the {\em average}
number of matrices
\begin{equation}
  \bar{N}_A= \avg{(q-1)^\Lambda \frac{\Lambda!}{\prod_i K_i! \prod_j C_j!}}{\bK,\bC,\Lambda}.
\end{equation}

Note that we can write the joint probability distribution as
\begin{equation}
  \prob{\bK,\bC,\Lambda} = \prob{\bK|\Lambda,\bC}\prob{\Lambda|\bC}\prob{\bC},
\end{equation}
and that $\prob{\Lambda|\bC}=\delta\prs{\Lambda,\sum_j C_j}$. Therefore, we have obtained
for the average number of matrices
\begin{equation}
  \bar{N}_A= \sum_\bK \sum_\bC \prob{\bK|\bC}\prob{\bC} (q-1)^{\sum_j C_j}
             \frac{\prs{\sum_j C_j}!}{\prod_i K_i! \prod_j C_j!},
\end{equation}
where the distribution $\prob{\bK|\bC}$ includes the constraint $\delta\prs{\sum_i K_i,\sum_j C_j}$.

A simple calculation shows that for the regular case, where all $C_j$'s and $K_i$'s  are fixed (to $C$ and $K$, respectively), and $q=2$,
the number of matrices scales as $N^{CN}$. Therefore, a more appropriate quantity to calculate instead of the average number
of matrices would be the quenched entropy
\begin{equation}
  \Xi \equiv \avg{\frac1N \ln N_A}{}
      =      \frac1N \sum_\bK \sum_\bC \prob{\bK|\bC}\prob{\bC} \ln \col{(q-1)^{\sum_j C_j}
             \frac{\prs{\sum_j C_j}!}{\prod_i K_i! \prod_j C_j!}},
\end{equation}
which scales as $\ln N$.

We analyze the behavior of this quantity for three different
cases. We choose each $C_j$ to be i.i.d. and $K$ to be chosen from
a multinomial distribution
\begin{equation}
  \prob{\bK}=\frac{(\sum_i K_i)!}{\prod_i K_i!} \frac1{N^{\sum_i K_i}} \, \delta\prs{\sum_i K_i, \sum_j C_j},
\end{equation}
for each realization of $\bC$. The three probability distributions
for the variables $C_j$ to be analyzed are
\begin{enumerate}
  \item uniform in the interval $[0,2\bar{C}]$
    \begin{equation}
      \prob{C_j} = 1/(2\bar{C}+1);
    \end{equation}
  \item binomial in the interval $[0,M]$
    \begin{equation}
      \prob{C_j} = \binom{M}{C_j} \prs{\frac{\bar{C}}{M}}^{C_j}\prs{1-\frac{\bar{C}}{M}}^{M-C_j};
    \end{equation}
  \item Zipf distribution for $C_j=1,...,M$
    \begin{equation}
      \prob{C_j} = \frac{C_j^{-s}}{\sum_{n=1}^K n^{-s}},
    \end{equation}
\end{enumerate}
where $\bar{C}$ is the mean of the distributions. The motivation for choosing these connectivity profiles
is that they appear to be the most commonly analyzed and feature (especially the latter) in recent
analysis and modeling of networks.

\begin{figure}
\centering
\includegraphics[width=14cm]{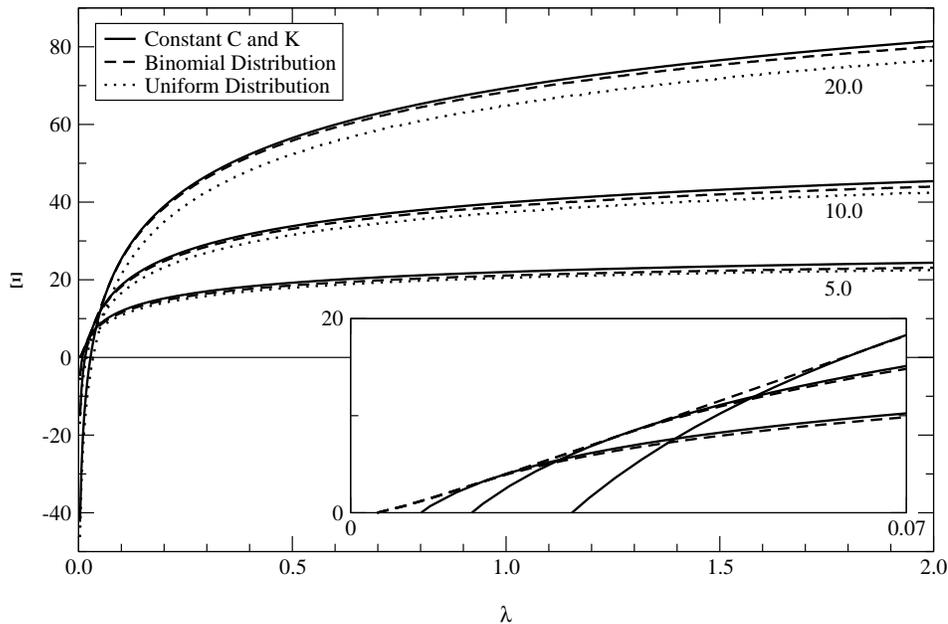}
\caption{Values of the quenched entropy $\Xi$ versus $\lambda$ for the different
         distributions and various $\bC$ values ($\bC=5,10,20$), with multinomial $\bK$:
         constant (continuous line),
         binomial (dashed line) and uniform (dotted line). The inset shows in detail the small $\lambda$ regime,
          where just the binomial and constant distributions are represented. The higher lines on the
	right correspond to the higher $\bC$ values. }
\label{figure:ANM}
\end{figure}

Results for the binomial (dashed line)
and uniform (dotted line) distributions with means
$\bar{C}=5.0,10.0,20.0$, $q=2$ and $N=300$ are plotted in Fig.~\ref{figure:ANM}, together
with the value of $\Xi$ with constant $C_j=\bar{C}$ and
$K_j=\bar{C}/\lambda$ values for all $i$ and $j$. This function is
explicitly given by
\begin{equation}
  \Xi^* = \bar{C}\ln (q-1) - \ln \bar{C}! +\frac1N \ln \prs{NC}! -\lambda \ln \prs{\bar{C}/\lambda}!,
\end{equation}
and we can obtain its asymptotic behavior for small and large
$\lambda$ as
\begin{align}
  \lambda \ll 1 \Rightarrow& \Xi^* = \bar{C} \ln (q-1) - \ln \bar{C}! +\bar{C}\ln \lambda N,\\
  \lambda \gg 1 \Rightarrow& \Xi^* = \bar{C} \ln (q-1) - \ln \bar{C}! +\bar{C}\ln \bar{C} N +(\gamma-1)\bar{C},
\end{align}
where $\gamma\approx 0.577216$ is the Euler-Mascheroni constant. Asymptotic limits for large $\lambda$ are given in
table~\ref{table:al}.

For large $\lambda$ values the result for constant $\bC$ and $\bK$ upper-bounds
the other two distributions. Additional calculations seem to indicate that it is always the case for any distribution, although a proof for this conjecture is still sought. This implies that if we keep the number of columns constant
and increase the ratio $\lambda$ by adding rows, whenever the number of rows is much larger than the number
of columns, the average number of matrices becomes independent of both the ratio and number of rows.
The plots also suggest that the average number of matrices in these cases are basically defined by the average value of
the $\bC$ distributions.

For small values of $\lambda$, the uniform distribution continues
to be upper-bounded by the constant distribution. The binomial
distribution, however, is higher for a small interval around
zero. This behavior is shown in the inset where lower $\bC$ values give rise
to higher $\Xi$ as $\lambda$ becomes smaller.

\begin{table}
  \centering
  \caption{Asymptotic values of $\Xi^*$ for large $\lambda$}
  \smallskip
  \begin{tabular}{|c|c|}
    \hline
    $\bar{C}$ & As. Value\\
    \hline \hline
    5  & 29.66 \\
    10 & 60.73 \\
    20 & 123.20 \\
    \hline
  \end{tabular}
  \label{table:al}
\end{table}

\begin{figure}
\centering
\includegraphics[width=12cm]{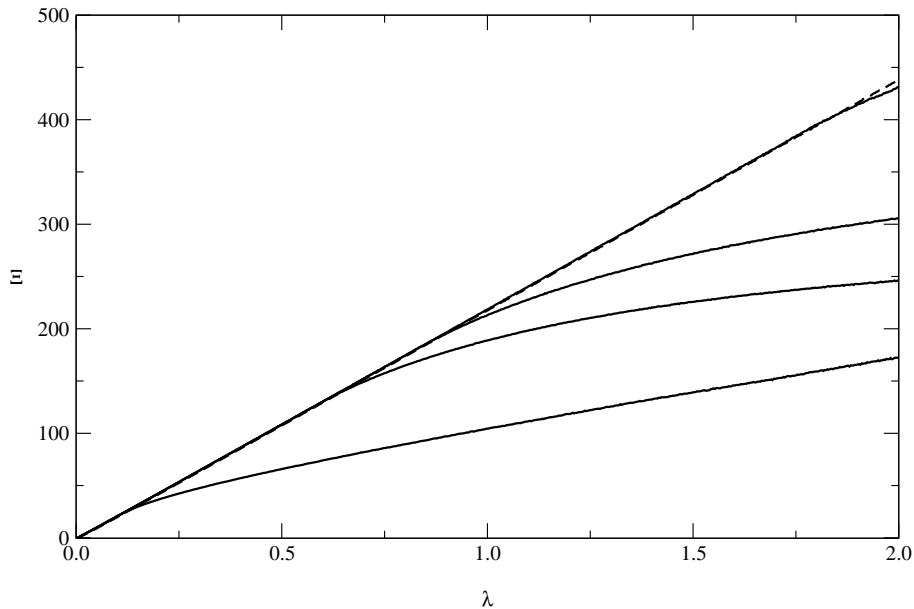}
\caption{Values of $\Xi$ versus $\lambda$ for the uniform distribution (dashed line) and the Zipf distribution
         (continuous lines) for $s=1,3,4,10$, respectively, from bottom to top.}
\label{figure:zipf}
\end{figure}

Figure~\ref{figure:zipf} shows the results for the Zipf distribution with different values for the power
$s$ compared with a uniform
distribution in the range $[0,M]$. In this case, the mean of the distributions vary with $\lambda$. We see that, although
the average value of the Zipf distributions increasingly differs from the uniform value $M/2$ as $s$ increases, the average
number of matrices actually becomes highly similar.

\section{Conclusions}
\label{section:Conclusions}

We have introduced a new mapping of Galois matrices to spin systems based on the group
homeomorphism between $GF(q)$ under
addition mod $q$ (denoted by $\oplus$) and the complex $q$-th roots of unity. In addition, we have introduced a different way for summing over random matrices
that can be generalized to include any kind of connectivity constraint and is
conceptually cleaner and simpler than the existing approaches.
 The new mapping and alternative summation over random matrices
allows for a factorization of the
constraints, which simplifies calculations of the kernel and the number of matrices
under various connectivity profiles.

Using the replica approach and these new introduced techniques, we
calculated the average dimension of the kernel for a general
distribution of non-zero entries and solved the resulting
equations numerically, finding that the average kernel density is
$1-M/N$ in all cases studied. We conjecture that this result is
always valid. Based on the analogy with thermodynamical quantities
corresponding to free energy, internal energy and Hamiltonian, we
showed that the replica symmetric ansatz in this case must be
exact. With the same techniques, we were also able to find the
total number of large matrices for fixed $\bK$ and $\bC$ and their
average number, which was then computed for different
 distributions of theoretical and practical relevance.

The
results presented have practical relevance in a number of areas, including
coding network modeling and some biological models.
With respect to LDPC codes, the average kernels density result
implies that randomly generated LDPC codes  typically define codes
of rate exactly $1-M/N$, an assumption which is generally made but
lacks rigorous derivations. Also, as the parity-parity check
matrix can represent the connectivities in graphs (see~\cite{Vicente00}),
the results obtained for the average number of matrices
provide a principled approach to determine the average number of possible
graphs with a given
connectivity distributions of a more general nature than the connectivity profiles examined in
this paper.

\section*{Acknowledgements}

Support from EPSRC grant EP/E049516/1 is gratefully acknowledged. R.C.A. would also like to thank 
Dr. Juan P. Neirotti for useful discussions.

\appendix

\section{Proof of $\Delta(q)=q$}
\label{appendix:Delta}

In this appendix we prove the statement made in
section~\ref{section:Kernel} that $\Delta(q)=q$ where
\begin{equation}
  \Delta(q) = \prod_{m=1}^{q-1} \col{1-\exp\prs{-\frac{2\pi i}{q} m}}.
\end{equation}

Let us use the notation
\begin{equation}
  z(m) \equiv e^{\frac{2\pi i}q m},
\end{equation}
and noting that unit complex roots appear in complex conjugate
pairs, we write
\begin{equation}
\label{eq:Delta}
   \Delta(q) =
      \left\{
      \begin{array}{cr}
        \prod_{m=1}^{(q-1)/2} \col{1-z(m)}\col{1-\bar{z}(m)}, & \mbox{$q$ odd},\\
        & \\
    2\prod_{m=1}^{(q-2)/2} \col{1-z(m)}\col{1-\bar{z}(m)}, & \mbox{$q$ even},
      \end{array}
      \right.
\end{equation}
where the bar indicates a complex conjugate. Using
\begin{equation}
  \begin{split}
    [1-z(m)][1-\bar{z}(m)] &= 2-2 \mbox{Re}[z(m)]\\
                           &= 2\col{1-\cos\prs{m\frac{2\pi}q}}\\
                           &= 4\sin^2\prs{m\frac{\pi}q},
  \end{split}
\end{equation}
equation~(\ref{eq:Delta}) becomes
\begin{equation}
   \Delta(q) =
      \left\{
      \begin{array}{cr}
        2^{q-1} \col{\prod_{m=1}^{(q-1)/2} \sin\prs{m\frac{\pi}q}}^2, & \mbox{$q$ odd},\\
        & \\
    2^{q-1} \col{\prod_{m=1}^{(q-2)/2} \sin\prs{m\frac{\pi}q}}^2, & \mbox{$q$ even}.
      \end{array}
      \right.
\end{equation}

As the $\sin$ function is positive in the interval $(0,\pi)$ and
$\sin(\pi/2)=1$ we can write, for any $q$,
\begin{equation}
   \label{equation:Delta}
   \Delta(q) = 2^{q-1} \prod_{m=1}^{q-1} \sin\prs{m\frac{\pi}q}.
\end{equation}

Using the known identity~\cite{Gradshteyn00}
\begin{equation}
  \sin (q x) = 2^{q-1} \prod_{m=0}^{q-1} \sin
  \prs{x+m\frac{\pi}q},
\end{equation}
divided by $\sin x$ and taking $x\rightarrow 0$, one obtains
\begin{equation}
  \prod_{m=1}^{q-1} \sin \prs{m\frac{\pi}q} =\frac{q}{2^{q-1}},
\end{equation}
which by substituting into equation~(\ref{equation:Delta}) gives
the desired result.

\section{Replica Symmetric Saddle Point Equations}
\label{appendix:Replicas}

Using integral representations for the first two sets of
Kroenecker delta functions, we can write the averaged replicated kernel
size defined in equation~(\ref{equation:parf}) as
\begin{equation}
  \begin{split}
    \parf &= \left<\frac1{\norm} \sum_\chs{\bv_a} \oint DWDZ \sum_\chs{A_{ij}}
             \col{\prod_{i,j} \prob{A_{ij}} \prs{W_i Z_j}^{\chi\prs{A_{ij}}}}\right.\\
          &  \quad\times\left. \prod_{i=1}^M \prod_a \delta\col{\bigoplus_{j=1}^N \prs{A_{ij}\otimes v_a^j},0}
             \right>_{\bK,\bC,\Lambda},
  \end{split}
\end{equation}
where $\otimes$ and $\oplus$ indicate multiplication and summation on $GF(q)$, respectively, and
\begin{equation}
  DWDZ=\col{\prod_{i=1}^M \frac{dW_i}{W_i^{K_i+1}}}\col{\prod_{j=1}^N\frac{dZ_j}{Z_j^{C_j+1}}}.
\end{equation}

Using the representation of the parity-check constraint given in equation~(\ref{equation:PCheck}),
the product over replica indices of the delta function can be written as
\begin{equation}
  \begin{split}
    \prod_a \delta\col{\bigoplus_{j=1}^N \prs{A_{ij}\otimes v_a^j},0}
      &= \prod_a \frac1q \prod_{m=1}^{q-1} \chs{1-\exp\prs{-\frac{2\pi i}{q} m}
         \prod_{j=1}^N \exp\col{\frac{2\pi i}{q} \prs{A_{ij}\otimes v_a^j}}}\\
      &= \frac1{q^n} \prod_a \col{1+\sum_{s=1}^{q-1} F_i(s,a) G(s)}\\
      &= \frac1{q^n} \sum_{r=0}^n \sum_\icn{a}{r} \sum_{s_1,\dots,s_r} G(s_1)\cdots G(s_r)
         F_i(s_1,a_1)\cdots F_i(s_r,a_r),
  \end{split}
\end{equation}
with
\begin{equation}
   G(s) \equiv \sum_\icn{m}{s} (-1)^s
               \exp\prs{-\frac{2\pi i}q m_1} \cdots \exp\prs{-\frac{2\pi i}q m_s},
\end{equation}
and
\begin{equation}
  \begin{split}
    F_i(s,a) &\equiv \exp\col{\frac{2\pi i}{q} \prs{A_{i1}\otimes v_a^1}}\cdots
                     \exp\col{\frac{2\pi i}{q} \prs{A_{iN}\otimes v_a^N}}\\
             &=      \prod_{j=1}^N \gamma_j(s,a,A_{ij}),
  \end{split}
\end{equation}
where we defined, for simplicity,
\begin{equation}
  \gamma_j(s,a,A_{ij}) \equiv \exp\col{\frac{2\pi i}q s \prs{A_{ij}\otimes v_a^j}}.
\end{equation}

We can now write the partition function as
\begin{equation}
  \parf = \avg{\frac1{\norm} \sum_\chs{\bv_a} \oint DZ \prod_{i=1}^M \frac1{q^n}
          \sum_{r=0}^n \sum_\icn{a}{r} \sum_{s_1,\dots,s_r} G(s_1)\cdots G(s_r)
          \oint \frac {dW_i}{2\pi i} \frac1{W_i^{K_i+1}} \Gamma_i }{\bK,\bC,\Lambda},
\end{equation}
where
\begin{equation}
  \begin{split}
    \Gamma_i &= \sum_{A_{i1},...,A_{iN}}
                \col{\prod_j \prob{A_{ij}} \prs{W_i Z_j}^{\chi\prs{A_{ij}}}}
                \prod_j \gamma_j(s_1,a_1,A_{ij})\cdots\gamma_j(s_r,a_r,A_{ij})\\
             &= \prod_j \sum_{A_{ij}}
                \prob{A_{ij}} \prs{W_i Z_j}^{\chi\prs{A_{ij}}}
                \gamma_j(s_1,a_1,A_{ij})\cdots\gamma_j(s_r,a_r,A_{ij})\\
             &= p^N \prod_j \col{1+\frac1{p} \sum_{h=1}^{q-1} \prob{A_{ij}=h} W_i Z_j
                \gamma_j(s_1,a_1,h)\cdots\gamma_j(s_r,a_r,h)},
  \end{split}
\end{equation}
where we define, for convenience, $p\equiv\prob{A_{ij}=0}$. Let us define a probability distribution over the values
of $h$ as
\begin{equation}
  \prob{h}=\frac{\prob{A_{ij}=h}}{1-p},
\end{equation}
in such a way that $h$ varies from 1 to $q-1$ and the probability over this range is correctly normalized. Then
\begin{equation}
  \begin{split}
    \Gamma_i &= p^N \prod_j \col{1+\prs{\frac{1-p}{p}} W_i Z_j \avg{\gamma_j(s_1,a_1,h)\cdots\gamma_j(s_r,a_r,h)}{h}}\\
             &= p^N \sum_{l=0}^N \sum_{\icn{j}{l}} \prs{\frac{1-p}{p}}^l W_i^l Z_{j_1}\cdots Z_{j_l}\\
             &  \quad\times
                \avg{\gamma_{j_1}(s_1,a_1,h)\cdots\gamma_{j_1}(s_r,a_r,h)}{h}\cdots
                \avg{\gamma_{j_l}(s_1,a_1,h)\cdots\gamma_{j_l}(s_r,a_r,h)}{h}.
  \end{split}
\end{equation}

The integrals over the $W_i$'s, acting on the $\Gamma_i$'s, select the power of $W_i$ to be $K_i$ and we therefore obtain
\begin{equation}
  \begin{split}
    \parf &=       \left<\kappa \sum_\chs{\bv_a} \oint DZ \prod_{i=1}^M \left\{
                   \sum_{r=0}^n \sum_\icn{a}{r} \sum_{s_1,\dots,s_r} G(s_1)\cdots G(s_r)
                   \sum_{\icn{j}{{K_i}}} Z_{j_1}\cdots Z_{j_{K_i}}\right.\right.\\
          &        \quad\times \left.\left.
                   \avg{\gamma_{j_1}(s_1,a_1,h)\cdots\gamma_{j_1}(s_r,a_r,h)}{h}\cdots
                   \avg{\gamma_{j_{K_i}}(s_1,a_1,h)\cdots\gamma_{j_{K_i}}(s_r,a_r,h)}{h}
                   \right\}\right>_{\bK,\bC,\Lambda}\\
          &\approx \left<\kappa \sum_\chs{\bv_a} \oint DZ \prod_{i=1}^M \left\{
                   \sum_{r=0}^n \sum_\icn{a}{r} \sum_{s_1,\dots,s_r} G(s_1)\cdots G(s_r)\right.\right.\\
          &        \quad\times\left.\left.\frac{N^{K_i}}{K_i!} \col{\frac1N\sum_{j=1}^N Z_j
                   \avg{\gamma_j(s_1,a_1,h)\cdots\gamma_j(s_r,a_r,h)}{h}}^{K_i}
                   \right\}\right>_{\bK,\bC,\Lambda}
  \end{split}
\end{equation}
where
\begin{equation}
  \kappa = p^{NM}\prs{\frac{1-p}{p}}^{\sum_i K_i} \norm^{-1} q^{-nM}.
\end{equation}

The calculation of $\norm$ is similar to the calculation of the number of matrices shown in appendix
\ref{appendix:Matrices} and
we end up with
\begin{equation}
  \kappa = \frac1{q^{nM} N^{(2)}_A},
\end{equation}
where $N^{(2)}_A$ is exactly the number of binary matrices ($q=2$) as calculated in appendix \ref{appendix:Matrices}.
Introducing the replica overlaps
\begin{equation}
  Q_\icn{a}{r}^{s_1,\dots,s_r} \equiv \frac1N \sum_{j=1}^N Z_j
                                      \avg{\gamma_j(s_1,a_1,h)\cdots\gamma_j(s_r,a_r,h)}{h},
\end{equation}
and the corresponding auxiliary variables $\hat{Q}_\icn{a}{r}^{s_1,\dots,s_r}$ by means of Dirac delta functions,
we can express the partition function as
\begin{equation}
  \begin{split}
    \parf &= \int DQD\hat{Q} \exp\prs{-N\sum Q_\icn{a}{r}^{s_1,\dots,s_r}\hat{Q}_\icn{a}{r}^{s_1,\dots,s_r}}\\
          &  \quad\times\left<\kappa \frac{N^{\sum_i K_i}}{\prod_i K_i!}
             \prod_i \col{\sum G(s_1)\cdots G(s_r) \prs{Q_\icn{a}{r}^{s_1,\dots,s_r}}^{K_i}}
             \right.\\
          &  \quad\times\left.\prod_j \chs{\sum_\chs{v_a^j} \oint DZ_j
             \exp\col{Z_j\sum \hat{Q}_\icn{a}{r}^{s_1,\dots,s_r}
             \avg{\gamma_j(s_1,a_1,h)\cdots\gamma_j(s_r,a_r,h)}{h}}}\right>_{\bK,\bC,\Lambda}\\
          &= \int DQD\hat{Q} \exp\prs{-N\sum Q_\icn{a}{r}^{s_1,\dots,s_r}\hat{Q}_\icn{a}{r}^{s_1,\dots,s_r}}\\
          &  \quad\times\left< q^{-nM} \frac{N^{\sum_i K_i}}{\prs{\sum_i K_i}!}
             \prod_i \col{\sum G(s_1)\cdots G(s_r) \prs{Q_\icn{a}{r}^{s_1,\dots,s_r}}^{K_i}}
             \right.\\
          &  \quad\times\left.\prod_j \chs{\sum_\chs{v_a^j} \col{\sum \hat{Q}_\icn{a}{r}^{s_1,\dots,s_r}
             \avg{\gamma_j(s_1,a_1,h)\cdots\gamma_j(s_r,a_r,h)}{h}}^{C_j}}\right>_{\bK,\bC,\Lambda}
  \end{split}
\end{equation}
where
\begin{equation}
  DQD\hat{Q}\equiv \prs{\prod \frac{dQ \, d\hat{Q}}{2\pi i/N}},
\end{equation}
and the summations run over all the allowed values of $r$, $\icn{a}{r}$ and $s_1,\dots s_r$.

Under the assumption of replica symmetry in the form
\begin{align}
  Q_\icn{a}{r}^{s_1,\dots,s_r}       &= Q_0 \avg{x^r}{x}, \\
  \hat{Q}_\icn{a}{r}^{s_1,\dots,s_r} &= \hat{Q}_0 \avg{\hx^r}{\hx},
\end{align}
where the averages over $x$ and $\hx$ are taken with respect to the field distributions $\pi(x)$ and $\hpi(\hx)$
respectively, we can show by straightforward algebraic manipulations that
\begin{align}
  \sum Q_\icn{a}{r}^{s_1,\dots,s_r} \hat{Q}_\icn{a}{r}^{s_1,\dots,s_r} &=
    Q_0\hat{Q}_0 \avg{\col{1+(q-1)x\hx}^n}{x,\hx},\\
  \sum G(s_1)\cdots G(s_r)\prs{Q_\icn{a}{r}^{s_1,\dots,s_r}}^{K_i} &=
    Q_0^{K_i} \avg{\chs{1+\col{\sum_s G(s)}\prod_{l=1}^{K_i} x_l}^n}{\bx},
\end{align}
where it is easy to see that
\begin{equation}
  \sum_s G(s) = \Delta(q)-1=q-1,
\end{equation}
and
\begin{equation}
  \begin{split}
    &\sum_\chs{v_a^j} \col{\sum \hat{Q}_\icn{a}{r}^{s_1,\dots,s_r}
      \avg{\gamma_j(s_1,a_1,h)\cdots\gamma_j(s_r,a_r,h)}{h}}^{C_j} =\\
      &\quad\hat{Q}_0^{C_j}\avg{\chs{\sum_{v=0}^{q-1} \prod_{l=1}^{C_j}\col{1+\omega(v,h_l)\hx_l}}^n}{\bhx,\bh},
  \end{split}
\end{equation}
with
\begin{equation}
  \omega(v,h_l) \equiv  \sum_{s=1}^{q-1} \exp\col{i \frac{2\pi s}q \prs{h_l\otimes v}}
                = \left \{
                    \begin{array}{cl}
                      q-1, & \text{if} \, h_l\otimes v=0,\\
                      -1,  & \text{otherwise}.
                    \end{array}
                  \right.
\end{equation}

We can simplify the last equation by noting that
\begin{equation}
  \sum_{v=0}^{q-1} \prod_{l=1}^{C_j}\col{1+\omega(v,h_l)\hx_l} =
    \prod_{l=1}^{C_j}\col{1+(q-1)\hx_l}+(q-1)\prod_{l=1}^{C_j}\prs{1-\hx_l}.
\end{equation}

Let us write
\begin{equation}
  \parf = \int DQD\hat{Q} \, e^{N\ts},
\end{equation}
with
\begin{equation}
  \ts = -\frac1N \ln N_A^{(2)} -n \lambda \ln q
        - Q_0\hat{Q}_0 \avg{\col{1+(q-1)x\hx}^n}{x,\hx} + \frac1N \ln \Phi,
\end{equation}
where
\begin{equation}
  \begin{split}
    \Phi &= \left<\frac{N^\Lambda}{\Lambda!} Q_0^\Lambda \hat{Q}_0^\Lambda
            \prod_i \avg{\col{1+(q-1)\prod_{l=1}^{K_i} x_l}^n}{\bx}\right.\\
         &  \quad\times\left.
            \prod_j \avg{\chs{\prod_{l=1}^{C_j}\col{1+(q-1)\hx_l}+(q-1)\prod_{l=1}^{C_j}\prs{1-\hx_l}}^n}{\bhx}
            \right>_{\bK,\bC,\Lambda}
  \end{split}
\end{equation}

Let us define $\alpha\equiv NQ_0\hat{Q}_0$. For $n\ll 1$, we can consider only the leading contributions in the number
of replicas, which gives
\begin{equation}
  \begin{split}
    \ln \Phi &=  \ln \epsilon(\alpha) + \frac{n}{\epsilon(\alpha)} \sum_i
                 \tavg{\frac{\alpha^\Lambda}{\Lambda!} \avg{\ln \col{1+(q-1)\prod_{l=1}^{K_i} x_l}}{\bx}}\\
             &   \quad\frac{n}{\epsilon(\alpha)} \sum_j
                 \tavg{\frac{\alpha^\Lambda}{\Lambda!}
                 \avg{\ln\chs{\prod_{l=1}^{C_j}\col{1+(q-1)\hx_l}+(q-1)\prod_{l=1}^{C_j}\prs{1-\hx_l}}}{\bhx}},
  \end{split}
\end{equation}
with
\begin{equation}
  \epsilon(\alpha) = \tavg{\frac{\alpha^\Lambda}{\Lambda!}}.
\end{equation}

Substituting the above formulas in $\ts$ for $n\rightarrow0$, the extremization with respect to $Q_0$, $\hat{Q}_0$,
$\pi(x)$ and $\hpi(\hx)$ leads to the saddle point equations (\ref{equation:S_Point_x}), (\ref{equation:S_Point_hx}) and
(\ref{equation:S_Point_Q}).

\section{Number of Matrices}
\label{appendix:Matrices}

Here we give the detailed calculation of the average number of $GF(q)$ $(M)\times N$ matrices for large $N$ and $N$.
Repeating the formula given in section \ref{section:NMatrices}, we have
\begin{equation}
    N_A = \sum_\chs{A_{ij}}
          \col{\prod_{i=1}^M\delta\prs{\sum_{j=1}^N\chi\prs{A_{ij}},K_i}}
          \col{\prod_{j=1}^N\delta\prs{\sum_{i=1}^M\chi\prs{A_{ij}},C_j}}.
\end{equation}
with $\chi\prs{A_{ij}}=0$ if $A_{ij}=0$ and 1 otherwise. Following a similar procedure as in \ref{appendix:Replicas}, we
use the integral representations of the Kroenecker delta functions to write it as
\begin{equation}
  \begin{split}
    N_A &= \oint DWDZ \prod_{i,j} \sum_{A_{ij}} (W_i Z_j)^{\chi\prs{A_{ij}}} \\
        &= \oint DWDZ \prod_{i,j} \col{ 1+(q-1)W_i Z_j} \\
        &= \oint DWDZ \prod_i \col{1+\sum_{r=1}^N (q-1)^r W_i^r \sum_\icn{j}{r} Z_{j_1}\cdots Z_{j_r}}\\
        &= \oint DWDZ \col{1+\sum_{s=1}^M \sum_\icn{i}{s} \sum_{r_1,...,r_s} (q-1)^{r_1+\cdots+r_s}
           W_{i_1}^{r_1}\cdots W_{i_s}^{r_s} F(r_1,Z)\cdots F(r_s,Z)},
  \end{split}
\end{equation}
where
\begin{equation}
  F(r,Z) \equiv \sum_\icn{j}{r} Z_{j_1}\cdots Z_{j_r}.
\end{equation}

The integrals over the $W$'s can pass through the summations and will factorize to give the corresponding Kroenecker
delta functions resulting in
\begin{equation}
  \begin{split}
    N_A &=       (q-1)^{\sum_i K_i} \oint DZ F(K_1,Z)\cdots F(K_M,Z)\\
        &=       (q-1)^\Lambda \oint DZ F(K_1,Z)\cdots F(K_M,Z)\\
        &=       (q-1)^\Lambda \oint DZ \prod_i \sum_\icn{j}{K_i} Z_{j_1}\cdots Z_{j_{K_i}}\\
        &\approx (q-1)^\Lambda \oint DZ \prod_i \frac1{K_i!} \prs{\sum_{j=1}^N Z_j}^{K_i}\\
        &=       (q-1)^\Lambda \oint DZ \frac1{\prod_i K_i!} \prs{\sum_{j=1}^N Z_j}^{\sum_i K_i}\\
        &=       \frac{(q-1)^\Lambda}{\prod_i K_i!} \oint DZ \sum_{j_1,...,j_\Lambda} Z_{j_1}\cdots Z_{j_\Lambda}\\
        &=       \frac{(q-1)^\Lambda}{\prod_i K_i!} \binom{\Lambda}{C_1}\binom{\Lambda-C_1}{C_2}\cdots
                 \binom{\Lambda-C_1-\cdots-C_{N-1}}{C_N},
  \end{split}
\end{equation}
which gives the final result
\begin{equation}
  N_A= \frac{(q-1)^\Lambda \Lambda!}{\prod_i K_i! \prod_j C_j!}.
\end{equation}

\section{Proof of Replica Symmetry}
\label{appendix:Proof_RS}

Using the fact that the random matrices can be seen as statistical physics systems with Hamiltonian
$\mathcal{H}(\bv) \equiv N-\ln \delta\prs{A\bv,0}$
we now prove that this implies that the replica symmetric solution
is the exact one. In fact, the form of the Hamiltonian implies that
\begin{equation}
  \prob{\bv} = \col{\sum_\bv \delta\prs{A\bv,0}}^{-1}=q^{-d(A)}.
\end{equation}

The distribution of the overlaps of the spins is given by
\begin{equation}
  \begin{split}
    \prob{\rho} &= \avg{\delta\prs{\rho-\frac1N \sum_{j=1}^N \sigma^j \sigma'^j}}{\gb{\sigma},\gb{\sigma}'}\\
                &= q^{-2d(A)} \sum_{\bv,\bv'} \delta\prs{A\bv,0} \delta\prs{A\bv',0}
                   \delta\col{\rho-\frac1N \sum_{j=1}^N \exp\prs{\frac{2\pi i}{q} \prs{v^j+v'^j}}}.
  \end{split}
\end{equation}

Let us call
\begin{equation}
  g(\bv,\bv') \equiv \delta\col{\rho-\frac1N \sum_{j=1}^N \exp\prs{\frac{2\pi i}{q} \prs{v^j+v'^j}}},
\end{equation}
and note that $g(\bv,\bv')=g(0,\bv\oplus\bv')$. Therefore we can write
\begin{equation}
  \begin{split}
    \prob{\rho} &= q^{-2d(A)} \sum_{\bv,\bv'} \delta\prs{A\bv,0} \delta\prs{A\bv',0} g(0,\bv\oplus\bv')\\
                &= q^{-2d(A)} \sum_{\bv,\bv'} \delta\prs{A\bv,0} \delta\prs{A\bv',0} \sum_{\bf u}
                   \delta\prs{{\bf u}, \bv\oplus\bv'} g(0,{\bf u})\\
                &= q^{-2d(A)} \sum_{\bf u} g(0,{\bf u})
                   \col{\sum_\bv \delta\prs{A\bv,0} \sum_{\bv'}  \delta\prs{A\bv',0}\delta\prs{{\bf u}, \bv\oplus\bv'}}\\
                &= q^{-2d(A)} \sum_{\bf u} g(0,{\bf u})
                   \col{\sum_\bv \delta\prs{A\bv,0} \delta\prs{A\prs{{\bf u}\oplus(-\bv)},0}}\\
                &= q^{-d(A)} \sum_{\bf u} \delta\prs{A{\bf u},0} g(0,{\bf u})\\
                &= \avg{\delta\prs{\rho-\frac1N\sum_{j=1}^N \sigma^j}}{\gb{\sigma}}.
  \end{split}
\end{equation}

Therefore, the distribution of the overlaps is the same as the distribution of the magnetization in the spin systems. This implies that there is no spin glass phase in the system and, therefore, no replica symmetry
breaking~\cite{Nishimori01}. The above calculation can also be viewed as a consequence of the \emph{gauge invariance}
of the Hamiltonian with respect to the transformation $\bv\rightarrow\bv\oplus\bv'$, where $A\bv'=0$, which leads
basically to the same calculation above.

\bibliographystyle{prsty}
\bibliography{statphysecc,basicbooks,randmatrix,inftheory}

\end{document}